\documentclass[preprintnumbers,article,amsmath,amssymb,floatfix,10pt,prd,twocolumn,superscriptaddress,nofootinbib]{revtex4-2}

\usepackage{bm}
\usepackage{xcolor}

\usepackage{newtxtext}
\usepackage{orcidlink}
\usepackage{graphicx}
\usepackage{subcaption}
\usepackage{multirow}
\usepackage{hyperref}
\hypersetup{colorlinks=true,linkcolor=red,urlcolor=blue,citecolor=red}

\def\jnl@style{\it}
\def\aaref@jnl#1{{\jnl@style#1}}

\def\aaref@jnl#1{{\jnl@style#1}}

\def\aj{\aaref@jnl{AJ}}                   
\def\apj{\aaref@jnl{ApJ}}                 
\def\apjl{\aaref@jnl{ApJ}}                
\def\apjs{\aaref@jnl{ApJS}}               
\def\apss{\aaref@jnl{Ap\&SS}}             
\def\aap{\aaref@jnl{A\&A}}                
\def\aapr{\aaref@jnl{A\&A~Rev.}}          
\def\aaps{\aaref@jnl{A\&AS}}              
\def\mnras{\aaref@jnl{Mon.~Not.~Roy.~Astron.~Soc.}}             
\def\prd{\aaref@jnl{Phys.~Rev.~D}}        
\def\prc{\aaref@jnl{Phys.~Rev.~C}}  
\def\prl{\aaref@jnl{Phys.~Rev.~Lett.}}    
\def\qjras{\aaref@jnl{QJRAS}}             
\def\skytel{\aaref@jnl{S\&T}}             
\def\ssr{\aaref@jnl{Space~Sci.~Rev.}}     
\def\zap{\aaref@jnl{ZAp}}                 
\def\nat{\aaref@jnl{Nature}}              
\def\aplett{\aaref@jnl{Astrophys.~Lett.}} 
\def\apspr{\aaref@jnl{Astrophys.~Space~Phys.~Res.}} 
\def\physrep{\aaref@jnl{Phys.~Rep.}}      
\def\physscr{\aaref@jnl{Phys.~Scr}}       
\def\commat{\aaref@jnl{Comm.~Math.~Phys.}}              
\def\science{\aaref@jnl{Science}}               
\def\cqg{\aaref@jnl{Classical Quant.~Grav.}}            
\def\jpcs{\aaref@jnl{JPCS}}                                     
\def\ijmpd{\aaref@jnl{Int.~J.~Mod.~Phys.~D}}                    
\def\grg{\aaref@jnl{Gen.~Relat.~Gravit.}}               
\def\rpp{\aaref@jnl{Rep.~Prog.~Phys.}}          
\def\npa{\aaref@jnl{Nucl.~Phys.~A}}        
\def\lrr{\aaref@jnl{Living Rev.~Rel.}}                   
\def\jcap{\aaref@jnl{J.~Cosmology Astropart.~Phys.}}    
\def\rmp{\aaref@jnl{Rev.~Mod.~Phys.}}   
\def\epjc{\aaref@jnl{Eur.~Phys.~J.~C}} 
\def\plb{\aaref@jnl{~Phy.~Lett.~B}} 
\def\mpla{\aaref@jnl{Mod.~Phy.~Lett.~A}} 
\def\arxiv{\aaref@jnl{arxiv.org}}

\allowdisplaybreaks[1]
\renewcommand{\arraystretch}{1.1}
\begin{document}
\color{black}       
\title{Bayesian and Machine-Learning Analyses of Nonminimal $f(Q)$ Gravity and $H_0$ Tension}

\author{Simran Arora\orcidlink{0000-0003-0326-8945}}
\email{arora.simran@yukawa.kyoto-u.ac.jp, dawrasimran27@gmail.com}
\affiliation{Center for Gravitational Physics and Quantum Information, Yukawa Institute for Theoretical Physics, Kyoto University, 606-8502, Kyoto, Japan.}

\author{Mridul Patel\orcidlink{0009-0000-9563-0926}}
\email{mridul.patel@student.rmit.edu.au}
\affiliation{School of Mathematical Sciences, RMIT University, 124 LaTrobe Street, Melbourne, VIC, Australia, 3000 }
%

\begin{abstract}

In this study, the cosmological implications of nonminimally coupled $f(Q)$ gravity are examined within the metric–affine formalism, in which the nonmetricity scalar $Q$ couples directly to the matter Lagrangian. Within the symmetric teleparallel framework, a representative $f(Q)$ model is constructed, and the corresponding background cosmological equations are derived. The analysis aims to test whether this geometric formulation yields more consistent realizations of nonminimal matter–geometry couplings. A comprehensive statistical MCMC analysis is performed using cosmic chronometers, DESI BAO DR2, and Type~Ia supernovae from the Pantheon+, DESY5, and Union3 samples and CMB. To complement the statistical study, we employ machine learning methods, such as linear regression, support vector regression (SVR), and random forest algorithms, to evaluate the predictive performance and robustness of the data. The results indicate that a partial alleviation of the $H_0$ tension can be achieved for a broad range of parameter choices. Nonetheless, $f(Q)$ gravity emerges as a promising and flexible framework for late-time cosmology, motivating further exploration of extended models consistent with all observations.\\

\end{abstract}

\maketitle

\section{Introduction}

Over the past three decades, the $\Lambda$ Cold Dark Matter ($\Lambda$CDM) model, widely regarded as the standard model of cosmology, has stood as the cornerstone of our understanding of the Universe. within the framework of general relativity, it provides an exceptionally successful description of cosmic evolution at both the background and perturbative levels. Its success has been firmly supported by a wealth of high-precision observations, including the temperature and polarization anisotropies of the Cosmic Microwave Background (CMB) \cite{Planck:2018vyg}, the Baryon Acoustic Oscillation (BAO) \cite{eBOSS:2020yzd,DESI:2024mwx,DESI:2025zgx}, type Ia supernovae (SNeIa) \cite{SupernovaSearchTeam:1998fmf,SupernovaCosmologyProject:1998vns}, constraints from SDSS galaxy clustering and
weak lensing \cite{DES:2021wwk,Heymans:2020gsg}. Together, these probes have reinforced the $\Lambda$CDM paradigm, offering a remarkably consistent picture of cosmic history across a broad range of cosmic epochs.

However, as observational precision has steadily improved over the past decade, subtle but persistent discrepancies, commonly referred to as cosmological tensions, have begun to emerge \cite{Weinberg:1988cp,Zlatev:1998tr,Joyce:2014kja}. The most prominent among these is the long-standing Hubble tension, concerning the measured values of the Hubble constant $H_0$, which may indicate new physics beyond the $\Lambda$CDM model. The so-called Hubble tension arises from the significant discrepancy between early- and late-universe determinations of the present-day cosmic expansion rate \cite{DiValentino:2020vvd,DiValentino:2021izs,Yang:2018euj}. The Planck Collaboration, assuming the $\Lambda$CDM model, reports a value of $H_0 = (67.4 \pm 0.5)\, \mathrm{km\,s^{-1}\,Mpc^{-1}}$, which stands in $\sim 4.4\sigma$ tension with the direct, distance-ladder measurement obtained by the SH0ES Collaboration in 2019, $H_0 = (73.04 \pm 1.04)\, \mathrm{km\,s^{-1}\,Mpc^{-1}}$. The latter result is based on Hubble Space Telescope observations of 70 long-period Cepheid variables in the Large Magellanic Cloud~\cite {Riess:2021jrx}. When combined with complementary probes, such as gravitational lensing and time-delay measurements, the deviation between the two determinations increases to a statistical significance of approximately $5.3\sigma$ \cite{H0LiCOW:2019pvv,Abdalla:2022yfr}. 

Motivated by these tensions, a wide range of theoretical efforts have explored the possibility that late-time cosmic acceleration might have a dynamical origin. Broadly, these attempts fall into two main categories: the first retains general relativity as the underlying gravitational framework while introducing additional dynamical components such as dark energy fields or fluids and the second constructs modified gravity theories in which general relativity emerges as a limiting case, typically through the inclusion of extra degrees of freedom capable of driving the observed acceleration.

A wide range of theoretical frameworks has been proposed to extend or modify general relativity. One of the most direct strategies involves starting from the Einstein-Hilbert action and supplementing it with additional geometric scalars or curvature invariants. This procedure gives rise to several well-known extensions, including $f(R)$ gravity~\cite{Sotiriou:2008rp,delaCruz-Dombriz:2006kob,Sotiriou:2006hs}, $f(R,T)$ gravity \cite{Harko:2011kv,Fisher:2019ekh,Alvarenga:2013syu}, $f(G)$ gravity~\cite{Nojiri:2005vv,DeFelice:2008wz} Lovelock gravity~\cite{Lovelock:1971yv}, and the broader family of Horndeski and Galileon scalar--tensor theories~\cite{Kobayashi:2019hrl}. An alternative route is based on the teleparallel formulation of gravity, where torsion replaces curvature as the mediator of gravitational interaction. Modifications within this setting lead to theories such as $f(T)$ gravity~\cite{Cai:2015emx,Paliathanasis:2016vsw}, its generalizations $f(T,T_G)$~\cite{Harko:2014aja} and $f(T,B)$~\cite{Franco:2020lxx,Bahamonde:2017ifa}, as well as scalar-torsion models~\cite{Hohmann:2018vle}.  

A conceptually distinct and more recent direction employs the nonmetricity, in which gravity is described by an affine connection that is both curvature-free and torsionless but does not preserve metric compatibility. This formulation, originally introduced in \cite{Nester:1998mp,Adak:2005cd}, naturally leads to the $f(Q)$ class of theories~\cite{Heisenberg:2023lru,BeltranJimenez:2017tkd}. The $f(Q)$ framework recovers general relativity as a particular case while maintaining second-order field equations, thereby avoiding higher-derivative instabilities. Owing to these advantages, it has become a promising arena for exploring new cosmological dynamics and potential deviations from standard theory \cite{Lazkoz:2019sjl,Mandal:2020buf,Zhao:2021zab,Sokoliuk:2023ccw}.

It is a fundamental result of differential geometry that a general affine connection can always be decomposed into three independent parts~\cite{Hehl:1994ue}, namely
\begin{equation}
\Gamma^{\lambda}{}_{\mu\nu}
= \hat{L}^{\lambda}{}_{\mu\nu} + K^{\lambda}{}_{\mu\nu} + L^{\lambda}{}_{\mu\nu},
\label{eq:connection_decomposition}
\end{equation}
where $\hat{L}^{\lambda}{}_{\mu\nu}$ denotes the Levi--Civita connection associated with the metric $g_{\mu\nu}$, defined by
\begin{equation}
\hat{L}^{\lambda}{}_{\mu\nu} \equiv 
\frac{1}{2} g^{\lambda\beta}
\left( \partial_{\mu} g_{\beta\nu}
+ \partial_{\nu} g_{\beta\mu}
- \partial_{\beta} g_{\mu\nu} \right).
\label{eq:levicivita}
\end{equation}
The second term, $K^{\lambda}{}_{\mu\nu}$, represents the contortion tensor
\begin{equation}
K^{\lambda}{}_{\mu\nu} \equiv 
\frac{1}{2} T^{\lambda}{}_{\mu\nu} + T_{(\mu}{}^{\lambda}{}_{\nu)},
\label{eq:contortion}
\end{equation}
constructed from the torsion tensor $T^{\lambda}{}_{\mu\nu} \equiv 
2\,\Gamma^{\lambda}{}_{[\mu\nu]}$. The final term, $L^{\lambda}{}_{\mu\nu}$, corresponds to the disformation tensor,
\begin{equation}
L^{\lambda}{}_{\mu\nu} \equiv 
\frac{1}{2} g^{\lambda\beta}
\left( - Q_{\mu\beta\nu}
- Q_{\nu\beta\mu}
+ Q_{\beta\mu\nu} \right),
\label{eq:disformation}
\end{equation}
which is defined in terms of the nonmetricity tensor as $Q_{\rho\mu\nu} \equiv \nabla_{\rho} g_{\mu\nu}$. 
Depending on which of the three fundamental tensors, curvature, torsion, or nonmetricity, is set to vanish, different geometrical frameworks emerge. When the nonmetricity tensor is set to zero, the resulting spacetime corresponds to Riemann-Cartan geometry, characterized by curvature and torsion. If, instead, the torsion tensor vanishes, one recovers the familiar torsion-free geometry. Setting the curvature tensor to zero leads to the teleparallel formulation of gravity. Moreover, if both nonmetricity and torsion are set to zero, the affine connection reduces to the Levi-Civita connection, yielding a Riemannian geometry. Conversely, when nonmetricity and curvature vanish while torsion remains, the connection becomes the Weitzenböck connection, defining a Weitzenböck geometry. Finally, if both curvature and torsion are set to zero, the connection becomes a symmetric teleparallel one, leading to the geometrical framework underlying $f(Q)$ gravity. For more review on this gravity, check Refs. \cite{Arora:2022mlo,DAmbrosio:2021zpm,Arora:2020met,Arora:2020tuk,Najera:2021afa,Yang:2021fjy}.

Nonminimal couplings involving functions of the Ricci scalar $R$ have been widely explored~\cite{Koivisto:2005yk,Bertolami:2007gv,Olmo:2014sra,Harko:2010mv} due to their rich phenomenology. However, because $R$ contains higher derivatives, such theories are generally interpreted as effective models that may encounter inconsistencies in certain regimes. These issues can be alleviated within the metric–affine formulation, where the field equations remain of second order. Motivated by this, it is natural to reconsider nonminimal matter couplings in the context of $Q$-gravity. Since the nonmetricity scalar $Q$ involves only first derivatives of the metric, a coupling of the type $f_2(Q)\mathcal{L}_m$ leads to second-order equations of motion \cite{Harko:2018gxr,Lu:2019hra,Hazarika:2024alm}. This geometric construction provides a promising route toward more consistent and universal realizations of nonminimal matter–geometry coupling theories.

In this work, we investigate the cosmological implications of nonminimally coupled $f(Q)$ gravity, with an emphasis on addressing the $H_0$ tension. A specific $f(Q)$ model is constructed and tested against observational data from multiple background probes. The analysis indicates that a partial alleviation of the $H_0$ discrepancy can be achieved within this framework across a broad range of parameter choices. To complement the theoretical analysis, machine learning techniques, such as linear regression, support vector regression (SVR), and random forest methods, are applied to evaluate the model's predictive performance and assess its consistency across different combinations of observational datasets.

The structure of this paper is as follows: Section~\ref{action}, provides a brief overview of $f(Q)$ gravity. Section~\ref{FLRW} presents the formulation of $f(Q)$ cosmology at the background level and introduces the specific model under consideration. The observational datasets, analysis, methodology, and statistical tools used for comparison are described in Section~\ref{data}. The main cosmological results are discussed in Section~\ref{results}, followed by the implementation of machine-learning techniques in Section~\ref{ML}. Finally, Section~\ref{conc} summarises the key findings and offers concluding remarks.

\section{General Framework} \label{action}

In this study, a gravitational action characterized by two arbitrary functions is considered, and can be written as \cite{Harko:2018gxr}  
\begin{equation}
S = \int d^4x \, \sqrt{-g} \left[ \frac{1}{2} f_1(Q) + f_2(Q)\, \mathcal{L}_M \right],
\label{eq:action}
\end{equation}
where $\mathcal{L}_M$ denotes the Lagrangian density of the matter fields, $f_1$ and $f_2$ are arbitrary functions of $Q$. The non-metricity conjugate comes with the combinations of the two independent traces $Q_\alpha = Q_{\alpha\;\;\mu}^{\;\;\mu}$ and  $\tilde{Q}_\alpha = Q^{\mu}{}_{\alpha\mu}$ which together characterize the deviation from metric compatibility. It is also convenient to introduce the corresponding superpotential tensor, 
\begin{align}
P^{\alpha}{}_{\mu\nu} &= -\tfrac{1}{2} L^{\alpha}{}_{\mu\nu} 
+ \tfrac{1}{4}\left(Q^\alpha - \tilde{Q}^\alpha\right) g_{\mu\nu} 
- \tfrac{1}{4}\delta^\alpha{}_{(\mu}\tilde{Q}_{\nu)}.
\label{eq:superpotential}
\end{align}
With this definition, the nonmetricity scalar takes the form
\begin{equation}
Q = - g^{\mu\nu} (L^{\alpha}{}_{\beta\mu} L^{\beta}{}_{\nu\alpha} 
- L^{\alpha}{}_{\beta\alpha} L^{\beta}{}_{\mu\nu}).
\label{eq:Q_scalar}
\end{equation} 
By construction, the nonmetricity formulation of gravity is equivalent to the Einstein-Hilbert Lagrangian, when the covariant derivative reduces to the ordinary partial derivative, i.e., \(\nabla_{\alpha} \circeq \partial_{\alpha}\). This particular gauge choice, indicated by the superscript $\circ$, is known as the coincident gauge \cite{BeltranJimenez:2017tkd}, and it has been shown to be a consistent and well-defined choice within the framework of symmetric teleparallel geometry.

The non metricity scalar is made of the contraction of the non-metricity tensor $Q_{\alpha \mu \nu}$, and its conjugate $P^{\alpha\mu\nu}$ as \cite{BeltranJimenez:2017tkd}
\begin{equation}
Q = - Q_{\alpha\mu\nu} P^{\alpha\mu\nu}.
\end{equation}
For clarity and compactness of notation, we introduce the following definitions,
\begin{equation}
f = f_1(Q) + 2 f_2(Q)\mathcal{L}_M, \quad 
F = f_1'(Q) + 2 f_2'(Q)\mathcal{L}_M,
\label{eq:fdefs}
\end{equation}
where a prime denotes differentiation with respect to \( Q \). Furthermore, we define the variations,
\begin{align}
&T_{\mu\nu} = -\frac{2}{\sqrt{-g}} 
\frac{\delta \big(\sqrt{-g}\,\mathcal{L}_M \big)}{\delta g^{\mu\nu}},
\label{eq:EMtensor} \\
&H_{\alpha}{}^{\mu\nu} = -\frac{1}{2} 
\frac{\delta \big(\sqrt{-g}\,\mathcal{L}_M\big)}{\delta \Gamma^{\alpha}{}_{\mu\nu}},
\label{eq:hyper}
\end{align}
which correspond to the energy-momentum tensor and the hypermomentum tensor density, respectively.

Varying the action in Eq. \eqref{eq:action} with respect to the metric yields the gravitational field equation,
\begin{eqnarray}
\nonumber
\frac{2}{\sqrt{-g}} \nabla_\alpha \left( \sqrt{-g} F P^{\alpha}{}_{\mu\nu} \right)
+ \frac{1}{2} g_{\mu\nu} f_1 \\ 
+ F \left( P_{\mu\alpha\beta} Q_{\nu}{}^{\alpha\beta}
- 2 Q_{\alpha\beta\mu} P^{\alpha\beta}{}_{\nu} \right)
= - f_2 T_{\mu\nu}.
\label{FE}
\end{eqnarray}
Variation of the same action with respect to the connection admits two possible approaches to impose the symmetric teleparallel condition. In the first, often referred to as the inertial variation, the connection is fixed to its pure-gauge form directly at the level of the action. Alternatively, one may retain a general connection and introduce suitable Lagrange multipliers to enforce the vanishing of curvature and torsion~\cite{BeltranJimenez:2018vdo}. In either formulation, the resulting connection field equation takes the form
\begin{equation}
\nabla_\mu \nabla_\nu \left( \sqrt{-g}\, F P^{\mu\nu}{}_{\alpha} 
- f_2 H_{\alpha}{}^{\mu\nu}\right)  = 0.
\label{eq:field_eq_connection}
\end{equation}

We may also express a relation more explicitly in terms of the divergence of the energy–momentum tensor as \cite{Harko:2018gxr}
\begin{eqnarray}
\nonumber
&D_{\mu} T^{\mu}{}_{\nu} 
+ \frac{2}{\sqrt{-g}} \nabla_{\alpha} \nabla_{\beta} H_{\nu}{}^{\alpha\beta} \nonumber = -\frac{2}{\sqrt{-g} f_2} 
\left[
(\nabla_{\alpha} \nabla_{\beta} f_2) H_{\nu}{}^{\alpha\beta} \right.\\
&\left. + 2 f_{2,(\alpha} \nabla_{\beta)} H_{\nu}{}^{\alpha\beta}
\right] - \left( T^{\mu}{}_{\nu} - \delta^{\mu}{}_{\nu} \mathcal{L}_M \right)
\nabla_{\mu} \log f_2.
\label{EMT}
\end{eqnarray}
The second and third terms on the right-hand side originate from the nonminimal couplings of the hypermomentum and energy–momentum tensors, respectively.  
The former contributes directly to the dynamics, while the latter introduces terms that remain second order in derivatives, provided the matter Lagrangian \(\mathcal{L}_M\) contains no higher-order terms, thus ensuring the consistency of the formulation. Eq.~\eqref{EMT} further reveals that the coupling between the nonmetricity scalar \(Q\) and the matter fields leads to the non-conservation of the energy–momentum tensor, indicating an exchange of energy and momentum between geometry and matter within this framework.

\section{Homogeneous and Isotropic Background}
\label{FLRW}
We now turn to cosmological applications of the considered framework. To this end, let us consider an isotropic, homogeneous, and spatially flat Universe, whose geometry is described by the line element
\begin{equation}
ds^2 = -N^2(t)dt^2 + a^2(t)\,\delta_{ij}\,dx^i dx^j ,
\label{eq:FRWmetric}
\end{equation}
where $a(t)$ denotes the cosmic scale factor and $N(t)$ is the lapse function introduced for generality. The time-reparametrization freedom of the theory is retained and without loss of generality, the choice $N=1$ is adopted whenever convenient. The expansion and dilation rates are defined as $H = \frac{\dot{a}}{a}$ and $T = \frac{\dot{N}}{N}$, respectively. Working in the coincident gauge, it is straightforward to obtain $Q = 6\left(\frac{H}{N}\right)^2$. 

We assume that the matter sector is described by a standard perfect fluid, whose energy--momentum tensor is given by
\begin{equation*}
T_{\mu \nu}=(\rho+p)u_{\mu}u_{\nu}+p,g_{\mu \nu},
\end{equation*}
where $\rho$ and $p$ denote the energy density and thermodynamic pressure of the fluid, respectively. The four-velocity $u^{\mu}=(1,0,0,0)$ is tangent to the fluid worldlines and satisfies the normalization condition $u^{\mu}u_{\mu}=-1$. Under these assumptions, the field Eq. \eqref{FE} reduce to the generalized Friedmann equations
\begin{align}
f_2 \rho &= \frac{f_1}{2} - \frac{6F}{N^2} H^2, 
\label{eq:friedmann1} \\
- f_2 p &= \frac{f_1}{2} - \frac{2}{N^2} \big( \dot{F} - F T \big) H 
+ \frac{F}{N^2}\big( \dot{H} + 3H^2 \big).
\label{eq:friedmann2}
\end{align}
It is straightforward to verify that in the general relativistic limit, obtained for $f_1 = -Q$, $f_2 = 1$, and consequently $F = -1$, Eqs.~\eqref{eq:friedmann1}--\eqref{eq:friedmann2} reduce to the standard Friedmann equations. The continuity equation for matter follows directly from these relations given by
\begin{equation}
\dot{\rho} + 3H(\rho + p) 
= -\frac{6 f_2'}{f_2 N^2} \, H \big( \dot{H} - HT \big) (\mathcal{L}_M + \rho).
\label{eq:continuity_general}
\end{equation}
In the minisuperspace corresponding to the metric~\eqref{eq:FRWmetric}, 
the matter Lagrangian density reduces to $\mathcal{L}_M = -\rho$. As a result, the standard continuity equation is recovered
\begin{equation}
\dot{\rho} + 3H(\rho + p) = 0,
\label{eq:continuity}
\end{equation}
which expresses the local conservation of the matter energy-momentum tensor. For detailed analysis, check Ref. \cite{Harko:2018gxr}.

In the following analysis, the choice $N=1$ is adopted, corresponding to the standard FLRW geometry. Under this condition, the relevant scalars simplify to
\begin{equation}
Q = 6H^2, 
\qquad 
T = 0.
\label{eq:QFRW_simplified}
\end{equation}

Consequently, the generalized Friedmann Eqs. \eqref{eq:friedmann1}--\eqref{eq:friedmann2} take the simplified form
\begin{align}
3H^2 &= \frac{f_2}{2F}\left(-\rho + \frac{f_1}{2f_2}\right), 
\label{eq:friedmann_FRW1} \\
\dot{H} + 3H^2 + \frac{\dot{F}}{F}H 
&= \frac{f_2}{2F}\left(p + \frac{f_1}{2f_2}\right).
\label{eq:friedmann_FRW2}
\end{align}
By eliminating the common term $3H^2$ in Eqs.~\eqref{eq:friedmann_FRW1} and~\eqref{eq:friedmann_FRW2}, an evolution equation for the Hubble parameter is obatined,
\begin{equation}
\dot{H} + \frac{\dot{F}}{F}H = \frac{f_2}{2F}\,(\rho + p).
\label{eq:H_evolution}
\end{equation}


Through the combination of Eqs.~\eqref{eq:friedmann_FRW2} and~\eqref{eq:H_evolution}, the modified dynamics can be expressed in a form that closely resembles the standard Friedmann equations of general relativity\footnote{In this representation, the gravitational field equations are $3H^2 = \rho_{\rm eff}$ and $2\dot{H} + 3H^2 = -\,p_{\rm eff}$.}. For this purpose, we introduce the effective energy density $\rho_{\rm eff}$ and effective pressure $p_{\rm eff}$ of the cosmic fluid, defined as
\begin{align}
\rho_{\rm eff} &= -\frac{f_2}{2F}
\left(\rho - \frac{f_1}{2f_2}\right),
\label{eq:rhoeff} \\
p_{\rm eff} &= \frac{2\dot{F}}{F}\,H
- \frac{f_2}{2F}
\left(\rho + 2p + \frac{f_1}{2f_2}\right).
\label{eq:peff}
\end{align}

An important quantity that characterizes the kinematical behavior of the cosmic expansion is the deceleration parameter, defined as $q = -\frac{\dot{H}}{H^2} - 1$, which can be derived from Eq.~\eqref{eq:H_evolution} as
\begin{equation}
q = \frac{\dot{F}}{F}\frac{1}{H} - \frac{f_2}{2H^2F}(\rho + p) - 1.
\label{eq:qexplicit}
\end{equation}
In order to describe the background dynamics and the transition between decelerated and accelerated expansion, it is useful to introduce the effective equation-of-state parameter $w$,
\begin{equation}
w = \frac{p_{\rm eff}}{\rho_{\rm eff}} 
= \frac{-4F\dot{H} + f_2\!\left(\rho + 2p + \tfrac{f_1}{2f_2}\right)}
{f_2\!\left(\rho - \tfrac{f_1}{2f_2}\right)} .
\label{eq:weff}
\end{equation}

To explore more general cosmological scenarios, it is necessary to specify the functional forms of $f_1(Q)$ and $f_2(Q)$. In this analysis, we consider both functions as a simple power-law dependence on the nonmetricity scalar $Q$, namely,
\begin{equation}
f_1(Q) = -Q + \alpha Q^{2}, 
\qquad 
f_2(Q) = 1+\beta Q,
\label{eq:f1f2_powerlaw}
\end{equation}
where $\alpha$ and $\beta$ are constant coupling parameters. Since the nonmetricity scalar satisfies $Q=6H^2$, it carries dimensions $[Q]=L^{-2}$. Consequently, dimensional consistency of the functions $f_1(Q)$ and $f_2(Q)$ requires both coupling parameters to have dimensions $[\alpha]=[\beta]=\text{[length]}^2$.

The function $F(Q)$ introduced in Eq.~\eqref{eq:fdefs} takes the form
\begin{align}
F(Q) &= (-1+2\alpha Q) - 2\beta \rho.
\label{eq:FQ_general}
\end{align}
Substituting the expressions for $f_1$, $f_2$, and $F$ into Eq.~\eqref{eq:friedmann_FRW1} at the present time makes one of the model parameters dependent
\begin{equation}
\beta = \frac{18 H_{0}^2 \alpha-(1-\Omega_{m0}) }{6 H_{0}^2 \Omega_{m0}}.
\label{eq:rho_powerlaw}
\end{equation}
Hence, the evolution equation for the Hubble parameter can be obtained as
\begin{equation}
3H^2 =
\frac{(1+6\beta H^2)
\left[
\frac{-6H^2+36\alpha H^4}{2(1+6\beta H^2)}
-3H_0^2\Omega_{m0}(1+z)^3
\right]}
{2\left(-1+12\alpha H^2-6\beta H_0^2\Omega_{m0}(1+z)^3\right)} .
\end{equation}
Here $H_0$ and $\Omega_{m0} = \Omega_{b0}+\Omega_{dm0}$ are the present time $(z=0)$  Hubble parameter and matter density, respectively. Now, our Hubble expression remains with the cosmological parameters along with an additional parameter $\alpha$.

One can also rewrite Eq.~\eqref{eq:H_evolution} as a one-dimensional autonomous dynamical system governing the cosmological evolution. Rewriting it in the form
\begin{equation}
\dot{H} \left(1+ \frac{24H^2(\alpha-\beta \rho_{Q})}{-1+2\alpha Q-2\beta \rho}\right)
=
\frac{(1+\beta Q)\gamma \rho}{2(-1+2\alpha Q)-2\beta \rho},
\end{equation}
one can clearly identify the role of the nonminimal matter-geometry coupling in modifying the cosmological dynamics. In particular, the quantity
\begin{equation}
F(Q)=(-1+2\alpha Q)-2\beta \rho,
\end{equation}
controls the deviation from the standard cosmological evolution. The factor multiplying $\dot H$ effectively modifies the Hubble friction term, while the right-hand side acts as an effective cosmological driving source. As the Universe evolves and the matter density decreases, the subleading contributions induced by the nonminimal coupling become increasingly important and can effectively generate a late-time accelerated expansion. 

It is convenient to introduce the dimensionless parameters $\tilde{\alpha}\equiv \alpha Q_0, \tilde{\beta}\equiv \beta Q_0$ where $Q_0=6H_0^2$ denotes the present-day value of the nonmetricity scalar. In the following, we express all results in terms of the dimensionless parameter $\tilde{\alpha}$ (and $\tilde{\beta}$, when relevant).

\section{Data and Methodology} \label{data}

In this section, we present the observational data to generate the posterior distribution of the full cosmological parameter space through the MCMC sampler. 
	
\begin{itemize}
\item \textbf{CC Data:} This data set comprises 30 model-independent measurements of the Hubble parameter, commonly known as Cosmic Chronometers (CC). It probes the expansion history by using massive, passively evolving galaxies with old stellar populations and minimal star formation, offering reliable estimates of \( H(z) \) across different redshifts~\cite{Vagnozzi:2020dfn,Jimenez:2001gg,Moresco:2015cya}.

\item \textbf{SNeIa Data:} Type Ia supernovae (SNeIa) are widely used as standard candles due to their relatively uniform intrinsic luminosity. This data set provides measurements of the apparent magnitude \( m_{b}(z) \), from which the luminosity distance \( D_{L}(z) \) is inferred via the magnitude-redshift relation
\begin{equation}
			\mu \equiv m-M = 5\log(D_L/\text{Mpc}) + 25 \  ,
		\end{equation}
		where, \(m\) denotes the apparent magnitude of the supernova and \(D_L\) is the luminosity distance: 	
		\begin{equation}
			D_L({z}) = c(1+{z}) \int_0^{{z}} \frac{dz'}{H(z')} \ ,
		\end{equation} 
assuming a flat FLRW metric, and \(c\) is the speed of light in km/s. The model parameters are constrained by minimizing the chi-square ($\chi^2$) likelihood, defined as:
		\begin{equation}
			-2 \ln (\mathcal{L}) = \chi^2 = \Delta D^{T} \mathcal{C}^{-1} \Delta D_j\ ,
		\end{equation}
		where $\Delta D = \mu_{ Obs} - \mu_{Model}$, $C^{-1}$ denotes the inverse combined statistical and systematic covariance matrix of the SNe sample. We use three different SNe datasets, including Pantheon+ \cite{Brout_2022}, DESY5 \cite{DES:2024jxu,DES:2024hip} and Union3 \cite{Rubin:2023jdq}.
        
\textbf{PP Data:} This data set refers to the Pantheon+ compilation, which includes 1550 spectroscopically confirmed Type Ia supernovae. The catalog provides 1770 data samples, from which we use the observational column corresponding to the non-SH0ES-calibrated apparent magnitude \( m_{\rm obs} \). We denote this subset as ``PP'' throughout our analysis.
		
\textbf{DESY5 Data:} This data set consists of Type Ia supernovae observations from the Dark Energy Survey five-year sample (DES-SN5YR), comprising 1829 distinct SNe. It includes 194 nearby SNe with redshift \( z < 0.1 \) and 1635 DES SNe. For our analysis, we compute the likelihood using the distance modulus \( \mu \) and the full covariance matrix provided in the data release. We have updated our supernova analysis to incorporate the DES-Dovekie release, which replaces the earlier DESY5 sample. The DES-Dovekie compilation includes 197 low-redshift SNe~Ia and 1623 DES likely SNe~Ia, with improved calibration and systematic control, making it approximately more robust than the previous dataset.

\textbf{Union3 Data:} The Union3 compilation, produced by the Supernova Cosmology Project \cite{Rubin:2023jdq}, provides one of the most extensive and homogeneous samples of Type~Ia supernovae to date. It extends the earlier Union and Union2 releases with improved photometric calibration, refined systematic corrections, and a uniformly processed sample of more than 1400 spectroscopically confirmed SNe Ia spanning the redshift range \( z \simeq 0.01\text{--}2.26 \). The dataset offers critical constraints on the cosmic expansion history and the properties of dark energy. Union3 combines observations from multiple surveys, including SNLS, SDSS, Pan-STARRS, CSP, and several low-redshift programs, along with space-based data from the Hubble Space Telescope (HST). All light curves were reprocessed using a common SALT2 pipeline to ensure photometric consistency.

\item \textbf{DESI BAO:} This data set consists of Baryon Acoustic Oscillation (BAO) measurements from the Dark Energy Spectroscopic Instrument (DESI) Data Release II~\cite{DESI:2025zgx}, which extends and improves upon the earlier DR1 results~\cite{DESI:2024mwx,DESI:2019jxc,Moon:2023jgl}. The key observables are the ratios $\{D_M/r_d$, $D_H/r_d$, $D_V/r_d \} $, where \( D_M \) is the comoving angular diameter distance, \( D_H \) the Hubble distance, \( D_V \) the spherically averaged BAO distance, and \( r_d \) the comoving sound horizon at the drag epoch~\cite{eBOSS:2020yzd,DESI:2024mwx}. For this analysis, we treat \(r_d\) as a free parameter. 

\item {\textbf{CMB Data:} This data set is derived by compressing the full Planck data set \cite{Planck:2018vyg} as reported in \cite{Chen:2018dbv}. The observables correspond to distance priors -- the acoustic scale \(l_A\) and shift parameter \(R\). These priors effectively encode information about the CMB temperature power spectrum. The acoustic scale characterizes the angular scale of the sound horizon, influencing peak spacing, while the shift parameter affects the line-of-sight direction, impacting peak heights. The shift parameter is defined as \cite{Elgaroy:2007bv}: 
	\begin{equation}
		R(z_*) \equiv \frac{D_A(z_*) \sqrt{\Omega_{m0} H_0^2}}{c}\ ,
	\end{equation}
	where \(z_* = 1089.92\) is the redshift at photon decoupling \cite{Planck:2018vyg}, and \(D_A\) is the angular diameter distance for a flat geometry given by:
	\begin{equation}
		D_A = c \int_{0}^{z} \frac{dz}{H(z)} \ .
	\end{equation}
	Here, \(\Omega_{m0}\) denotes the total matter density at present. For the minimally coupled case, where both baryonic and dark matter are pressureless and individually conserved, baryon density does not explicitly appear in the Hubble parameter \(H = H_0 \sqrt{\Omega_\Lambda + \Omega_{m0} (1+z)^3 + \Omega_r (1+z)^4}\). However, the total matter density is \(\Omega_{m0} = \Omega_{dm0} + \Omega_{b0}\), since the non-gravitational coupling only modifies the dark matter profile while baryons remain pressureless with \(\Omega_b \propto (1+z)^3\). Therefore, care must be taken when applying such models. The acoustic scale is defined as:
	\begin{equation}
		l_A = \frac{\pi D_A(z_*)}{r_s(z_*)}
	\end{equation}
	where \(r_s\) is the comoving sound horizon at the photon decoupling epoch. Along with these, the compressed CMB data includes \(\Omega_{b_0} h^2\). The data set includes the correlation matrix between \(\{R, l_A, \Omega_{b_0} h^2\}\), whose inverse covariance matrix is reported in the appendix of \cite{Chen:2018dbv}. The chi-square is calculated as:
	\begin{equation}
		\chi^2_{\rm PLA} = \sum \rm (D_i^{ obs}- D_i^{th}) C_{ij}^{-1} (D_j^{ obs}- D_j^{th}) \ .
	\end{equation}
 In the current study, we use the compressed dataset, which has been further improved in \cite{Arendse:2019hev,Cai:2014ela,Liu:2024fjy} by including an additional observable. The sample is extended to a four-parameter likelihood, $\{100,\Omega_b h^2, 100,\theta_{*}, R, \Omega_{\rm dm} h^2\}$, along with its correlation matrix. This extension has been shown to be efficient in constraining models beyond $\Lambda$CDM. The current likelihood replaces the acoustic scale with the angular scale defined as $\theta_{*} = r_s(z_*)/D_A(z_*)$ and introduces an additional likelihood parameter corresponding to the dark matter density $\Omega_{\rm dm}$. Additionally, we compute the redshift at recombination and at the photon–baryon decoupling epoch using the Hu–Sugiyama fitting formula \cite{Hu:1995en}}
	\begin{eqnarray*}
	 z_{\mathrm{d}}&=&1345 \frac{\left.\left(\Omega_{\mathrm{m}} h^2\right)^{0.251}\left[1+b_1\left(\Omega_{\mathrm{b}} h^2\right)^{b_2}\right)\right]}{1+0.659\left(\Omega_{\mathrm{m}} h^2\right)^{0.828}}\ , \\
		b_1 &=&0.313\left(\Omega_{\mathrm{m}} h^2\right)^{-0.419}\left[1+0.607\left(\Omega_{\mathrm{m}} h^2\right)^{0.674}\right] \ , \\
	 b_2&=&0.238\left(\Omega_{\mathrm{m}} h^2\right)^{0.223} \ .
		\end{eqnarray*}
	and 
\begin{eqnarray*}
	z_*&=&1047\left[1+0.00124\left(\Omega_{\mathrm{b}} h^2\right)^{-0.738}\right]\left[1+g_1\left(\Omega_{\mathrm{m}} h^2\right)^{g_2}\right] \\
	g_1&=&0.0783\left(\Omega_{\mathrm{b}} h^2\right)^{-0.238}\left[1+39.5\left(\Omega_{\mathrm{b}} h^2\right)^{0.763}\right]^{-1} \\
	g_2&=&0.56\left[1+21.1\left(\Omega_{\mathrm{b}} h^2\right)^{1.81}\right] .
\end{eqnarray*}
\end{itemize}
We perform our analysis using probes: CC, DESI BAO DR2, and Type Ia supernovae (SNe) and CMB. We consider three combinations of observational data sets: \text{CMB + CC + DESI + PP}, \text{CMB + CC + DESI + DESY5}, and \text{CMB + CC + DESI + Union3}. The total likelihood is constructed as
\begin{eqnarray}
  -2\ln \mathcal{L} =  \chi^2_{\text{CMB}}+ \chi^2_{\text{CC}} + \chi^2_{\text{BAO}} +\chi^2_{\text{SNe}},  
\end{eqnarray}
and is used to constrain the model characterized by the free parameters \(\{\alpha, \gamma, H_0, \Omega_bh^2,\Omega_{dm}h^2\}\). To explore the parameter space and obtain best-fit values via \( \chi^2 \) minimization, we employ the \texttt{emcee} MCMC sampler~\cite{Foreman-Mackey:2012any}. For comparison, we also analyze the standard \( \Lambda \)CDM model with free parameters: \( \{\Omega_m, \Omega_bh^2, H_0\} \). The prior ranges for the free parameters are defined as follows:  $\alpha \in \mathcal{U}[0, 0.001]$, $\gamma \in \mathcal{U}[0, 1]$, $H_0 \in \mathcal{U}[30, 100]$~$\text{km\,s}^{-1}\,\text{Mpc}^{-1}$, $\Omega_bh^2 \in \mathcal{U}[0.01, 0.1]$, $\Omega_{dm}h^2 \in \mathcal{U}[0.01, 0.5]$. Posterior distributions are analyzed and visualized using the \texttt{GetDist} package~\cite{Lewis:2019xzd}. To assess the statistical performance of the models relative to flat \( \Lambda \)CDM, we employ two standard model selection criteria: the Akaike Information Criterion (AIC) and the Bayesian Information Criterion (BIC)~\cite{Akaike:1974vps,Schwarz:1978tpv}, defined as:
\begin{align}
    \mathrm{AIC} &= -2 \ln \mathcal{L}_{\text{max}} + 2k, \label{AIC} \\
    \mathrm{BIC} &= -2 \ln \mathcal{L}_{\text{max}} + k \ln N, \label{BIC}
\end{align}
where \( k \) is the number of free parameters, \( N \) is the total number of data points, and \( \mathcal{L}_{\text{max}} \) is the maximum likelihood. The relative model performance is quantified through $\Delta\mathrm{IC} = \mathrm{IC}_{\text{model}} - \mathrm{IC}_{\Lambda\mathrm{CDM}}$. Smaller values of $\Delta\mathrm{IC}$ indicate stronger statistical support for the corresponding model relative to the reference $\Lambda$CDM cosmology.

\begin{table*}
		\begin{tabular} { l  c c c c c} 
			\noalign{\vskip 3pt}\hline\noalign{\vskip 1.5pt}\hline\noalign{\vskip 5pt}
			\multicolumn{1}{c}{} &  \multicolumn{1}{c}{\bf CMB+DESI+CC+PP} & \multicolumn{1}{c}{\bf CMB+DESI+CC+DESY} &  \multicolumn{1}{c}{\bf CMB+DESI+CC+Union3} \\
			\noalign{\vskip 3pt}\cline{1-6}\noalign{\vskip 3pt}	
			Parameters &  68\% limits &  68\% limits &  68\% limits &\\ \hline	 
            \multicolumn{1}{c}{} &  \multicolumn{1}{c}{\bf } &  \multicolumn{1}{c}{\bf Model} & \multicolumn{1}{c}{} &  \multicolumn{1}{c}{} \\ \hline
			{\boldmath$ \tilde{\alpha}$} &   $0.598^{+0.0048}_{-0.0046}$ & $0.596^{+0.0044}_{-0.0045}$ & $0.597^{+0.0043}_{-0.0044}$ \\
		
			{\boldmath$\Omega_{dm}h^2$} & $0.2482 \pm {0.0033}$ & $0.2468 \pm {0.0032}$ & $0.2472 \pm {0.0031}$   \\ [0.5ex]

            {\boldmath$\Omega_{b}h^2$} & $0.02203 \pm {0.00021}$ & $0.02205 \pm {0.00021}$ & $0.02209 \pm {0.00020}$  \\ [0.5ex]
            
			{\boldmath $H_0 $} & $68.42 \pm 0.83$ & $68.67 \pm 0.85$ & $68.84 \pm 0.78$ &  \\ [0.5ex]
			
			{\boldmath$\gamma$} & $0.1767 \pm {0.0003}$ & $0.1767 \pm {0.0003}$ & $0.1767 \pm {0.0003}$ \\ [0.5ex]			
			\hline			
			\multicolumn{1}{c}{} &  \multicolumn{3}{c}{\bf $\Lambda$CDM} &  \multicolumn{1}{c}{}  &  \multicolumn{1}{c}{} \\
			\hline   \\[-1.5ex] \rule{0pt}{2ex}  
			{\boldmath$ \Omega_m$} & $0.301 \pm {+0.0038}$  &  $0.299 \pm {0.0037}$ & $ 0.299 \pm 0.0038$ & \\

             {\boldmath$\Omega_{b}h^2$} & $0.02256 \pm {0.00013}$ & $0.02258 \pm {0.00012}$ & $0.02258^{+0.00012}_{-0.00014}$  \\ [0.5ex]
			
			{\boldmath $H_0 $} & $68.43 \pm 0.29$ & $68.58 \pm 0.29$ & $68.53 \pm 0.30$ \\[0.5ex]
			\hline
{\boldmath $\Delta$AIC} & $-0.51$ & $7.04$ & $-0.27$ &   \\ [0.5ex]
{\boldmath $\Delta$BIC} & $10.11$ & $18.11$ & $4.04$ & \\ [0.5ex]
{\boldmath $logZ(model)$} & $-826.4$ & $-883.2$ & $-49.2$ & \\ [0.5ex]
{\boldmath $logZ(\Lambda CDM)$} & $-753.68$ & $-865.4$ & $-49.05$ & \\ [0.5ex]
{\boldmath $\chi_{red}^2$} & $0.97$ & $0.903$ & $0.911$ & \\ [0.5ex]
             \hline  \hline
		\end{tabular}
		\caption{Cosmological constraints on the model based on the CMB+DESI+CC+PP, CMB+DESI+CC+DESY, and CMB+DESI+CC+Union3. }
		\label{tab:best_fit}
	\end{table*}

\begin{figure}
    \centering
    \includegraphics[width=1.0\linewidth]{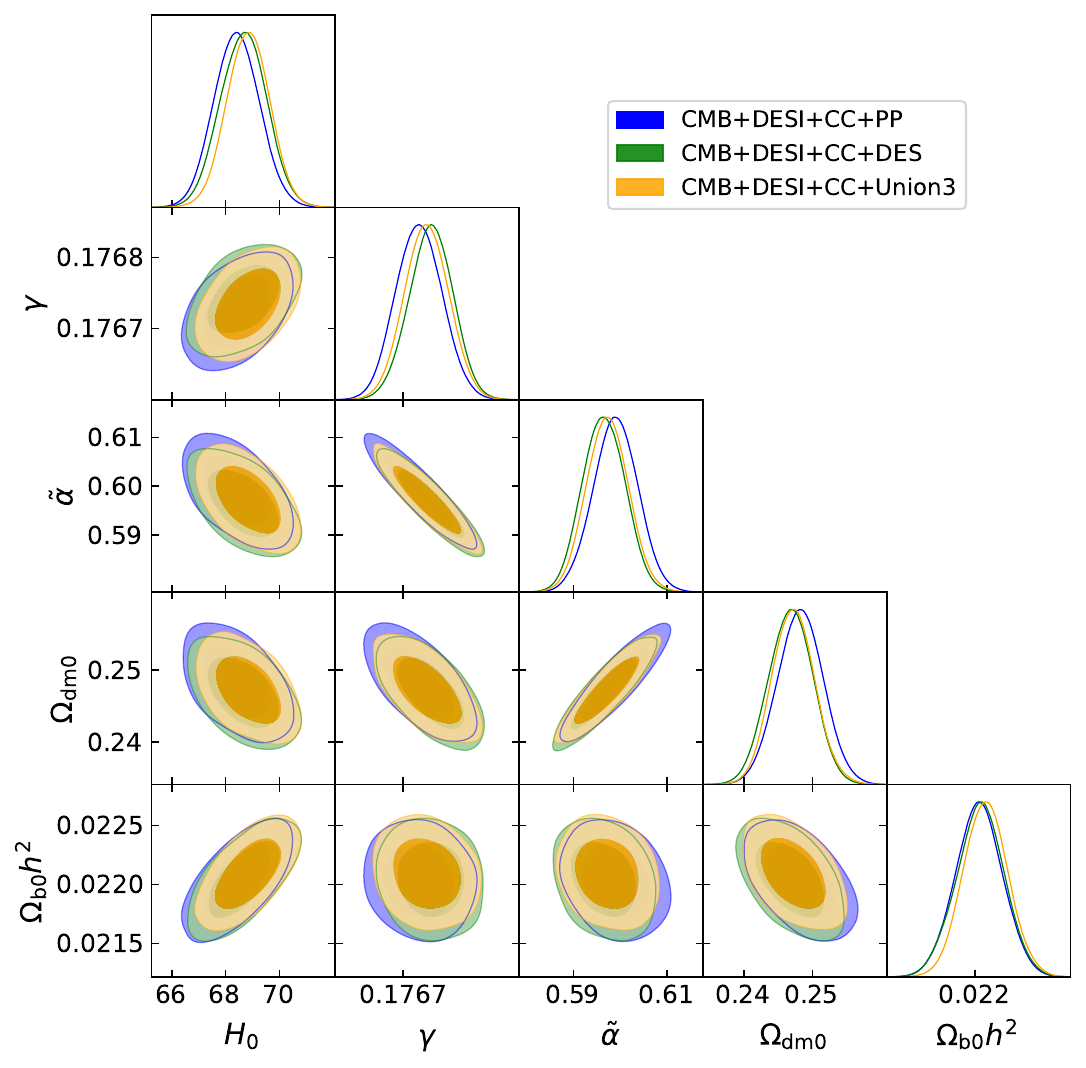}
    \caption{Two-dimensional contours of the parameter space for the $f(Q)$ model using different observational data. Here, $\tilde{\alpha}$ denotes the dimensionless coupling parameter corresponding to the original dimensional parameter $\alpha$.}
    \label{fig:con}
\end{figure}

\section{Results and Comparison} \label{results}

In this section, we present the observational constraints and reconstructed expansion history for the nonminimal $f(Q)$ gravity model using a combination of the observational data considered in the previous section.

Table~\ref{tab:best_fit} summarizes the 68\% confidence limits on the model parameters obtained from different dataset combinations, alongside the corresponding $\Lambda$CDM results for comparison. The parameters are tightly constrained across all combinations, with $\alpha$ showing small but consistent variations that remain well within statistical uncertainties. The Hubble constant lies in the range $H_0 \simeq 68$--$69~\mathrm{km\,s^{-1}\,Mpc^{-1}}$, and intermediate between the CMB and local distance-ladder determinations, suggesting a mild amelioration of the $H_0$ tension. The values of $\Delta$AIC, and $\Delta$BIC$ $ indicate that the nonminimal $f(Q)$ model provides a fit of comparable statistical quality to $\Lambda$CDM, while offering additional flexibility to describe the late-time expansion. All combinations yield reduced $\chi^2$ values close to unity, confirming the goodness of fit and internal consistency of the constraints. To further assess the statistical performance of the model relative to the standard $\Lambda$CDM cosmology, we also consider the Bayesian evidence through the Jeffreys’ scale \cite{Trotta:2008qt,Kass:1995loi}. In particular, we evaluate the quantity
\begin{equation}
\Delta \ln \mathcal{Z} = \ln \mathcal{Z}_{\rm model} -
\ln \mathcal{Z}_{\Lambda{\rm CDM}},
\end{equation}
where $\mathcal{Z}$ denotes the Bayesian evidence. According to the Jeffreys’ criterion, values $|\Delta \ln \mathcal{Z}|<1$ correspond to inconclusive evidence, while $1<|\Delta \ln \mathcal{Z}|<2.5$ indicates weak evidence, and larger values correspond to moderate or strong statistical preference. Although the proposed framework can reproduce the late-time accelerated expansion, the Bayesian evidence suggests that the additional model complexity is not sufficiently supported by the current observational data.

Fig.~\ref{fig:con} displays the marginalized posterior distributions of the parameters obtained from various combinations of observational data. The diagonal panels correspond to the one-dimensional posterior distributions, while the off-diagonal panels show the 68\% and 95\% confidence contours. Including the Union3, DESY, and PP samples significantly tightens the contours and reduces degeneracies, especially in the $(\alpha,\gamma)$.

\begin{figure}
    \centering
    \includegraphics[width=0.8\linewidth]{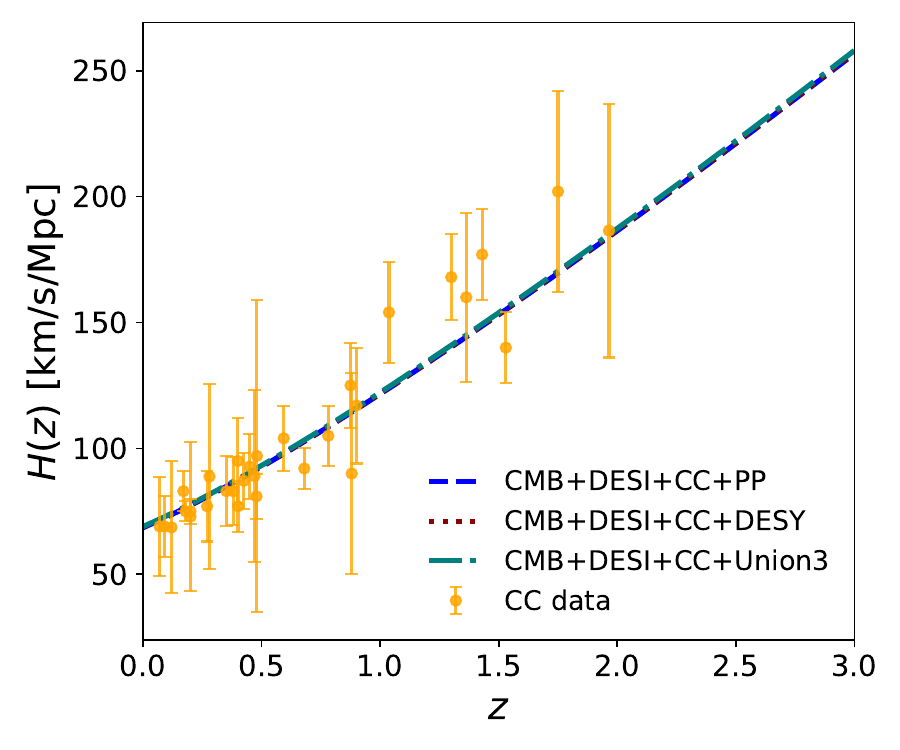} \\
        \includegraphics[width=0.8\linewidth]{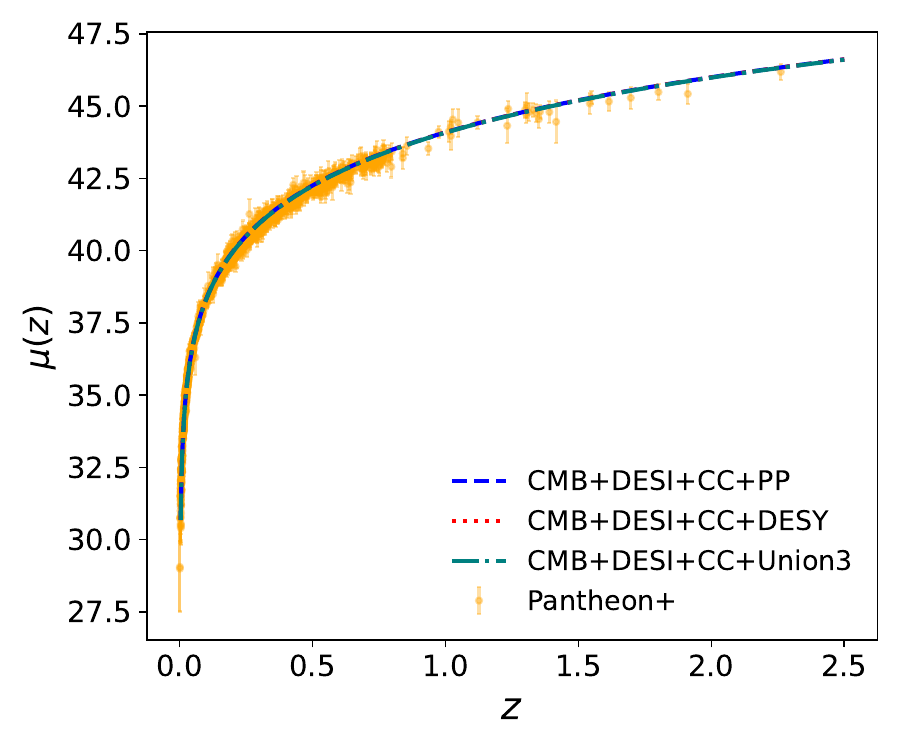}
    \caption{The behavior of cosmological parameters using best-fit parameters from observational data. The top panel shows $H(z)$ compared with cosmic chronometer data, and the bottom panel shows the distance modulus $\mu(z)$ against Pantheon$+$ supernova samples.}
    \label{fig:hz}
\end{figure}

The corresponding reconstructed expansion history is summarized in Fig.~\ref{fig:hz}. The upper panel compares the theoretical $H(z)$ predictions with the latest CC measurements, while the lower panel presents the distance modulus $\mu(z)$ as a function of redshift. The model reproduces both the observed expansion rate and luminosity-distance relations with high accuracy, remaining consistent with $\Lambda$CDM within the $1\sigma$ confidence region. 

\begin{figure*}
    \centering
    \includegraphics[scale=0.4]{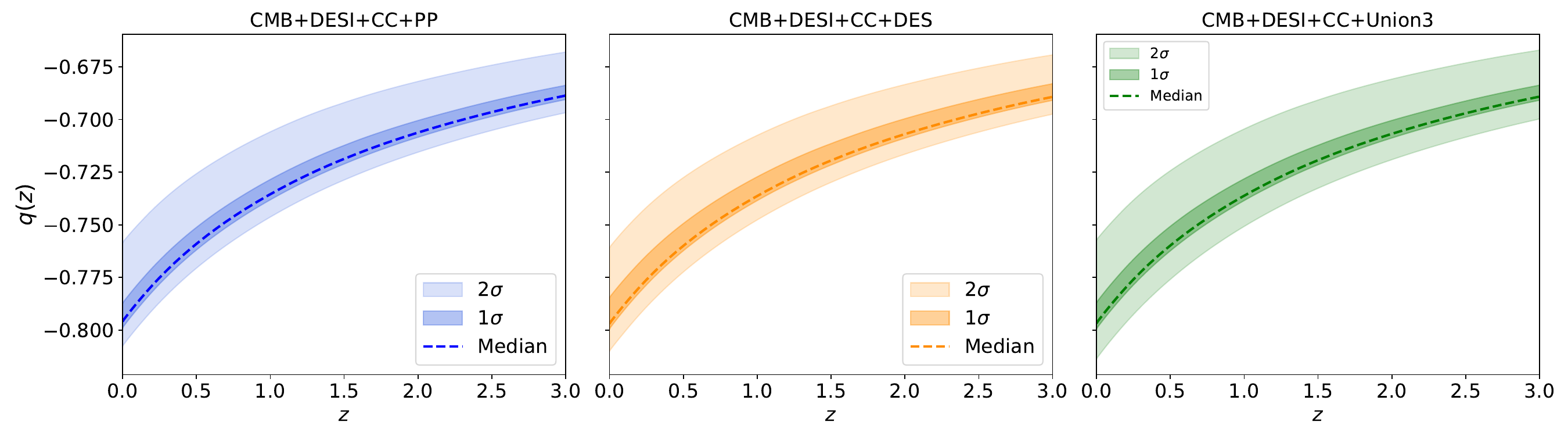}
     \includegraphics[scale=0.4]{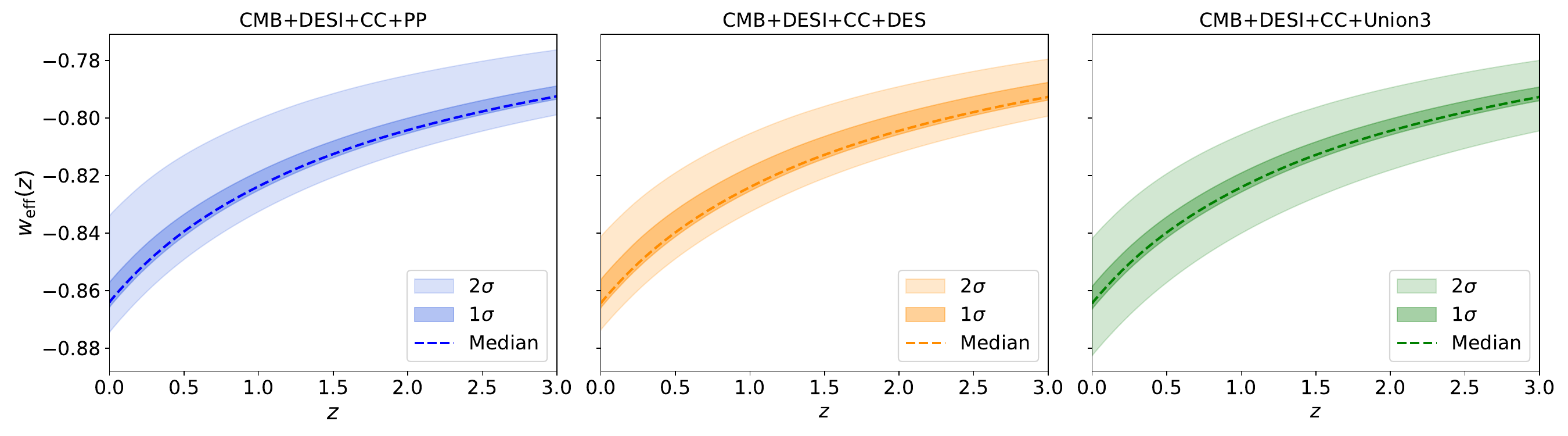}
    \caption{Evolution of the deceleration parameter $q(z)$ and equation of state $w(z)$ for the combinations of \text{CMB+DESI+CC+PP}, \text{CMB+DESI+CC+DESY}, and \text{CMB+DESI+CC+Union3}, respectively.}
    \label{fig:wz}
\end{figure*}

Further insight into the model’s dynamical behavior is provided by the evolution of the cosmographic parameters shown in Fig.~\ref{fig:wz}. The reconstructed cosmological evolution indicates that the Universe remains in a persistent accelerating phase throughout the considered redshift range. In particular, the deceleration parameter $q(z)$ retains negative values, typically around $q\sim -0.8$ at low redshift, providing clear evidence for late-time cosmic acceleration. Although the magnitude of acceleration gradually decreases toward higher redshift, the evolution remains entirely within the accelerated regime.

Simultaneously, the effective equation of state stays close to the cosmological constant boundary, with $w_{eff}(z)\sim -0.9$, indicating a dark-energy-dominated expansion history that closely mimics a $\Lambda$CDM-like behavior. The smooth monotonic evolution of both the parameters, together with the consistency across different supernova datasets, suggests that the model provides a stable and observationally viable description of the late-time Universe.


\begin{figure*}
    \centering
    \includegraphics[width=1\linewidth]{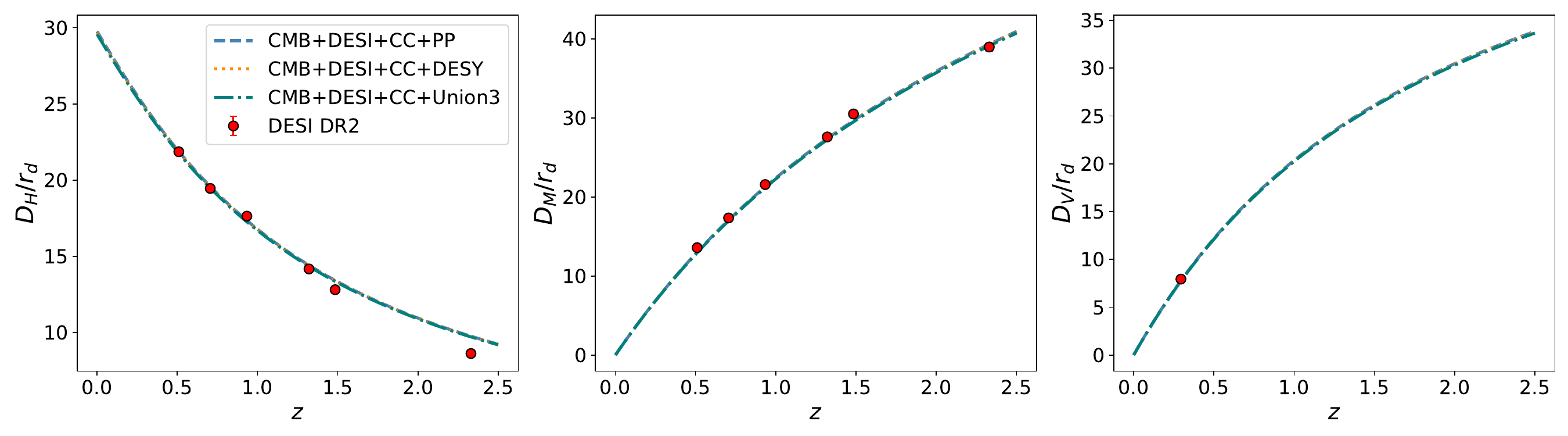}
    \caption{The panel displays the cosmological parameters plotted alongside DESI BAO data. The first plot shows the behavior of $D_H/rd$, the second presents the $D_M/rd$, and the last illustrates the $D_V/rd$ profile.}
    \label{fig:bao}
\end{figure*}

Finally, the comparison with BAO distance measurements from DESI\,DR2 is shown in Fig.~\ref{fig:bao}. The three panels correspond to the normalized cosmological distance indicators $D_H/r_d$, $D_M/r_d$, and $D_V/r_d$, representing the Hubble distance, transverse comoving distance, 
and volume-averaged distance, respectively. These quantities are directly constrained by BAO observations and provide a geometric cross-check of the model. The theoretical predictions from all dataset combinations (CMB+CC+BAO+PP$+$, CMB+CC+BAO+Union3, and CMB+CC+BAO+DESY) show good agreement with the DESI\,DR2 data, maintaining the overall consistency across all three panels underscores the stability of the model. 



Taken together, the results are mutually consistent across all observational probes, reinforcing the viability of the nonminimal $f(Q)$ model and demonstrating that DESI\,BAO\,DR2 is already making a significant contribution to advancing precision cosmology early in its operational timeline. Next, we discuss the physical interpretation of these findings in the context of current cosmological tensions. 

\begin{figure*}
    \centering
    \includegraphics[width=0.6\linewidth]{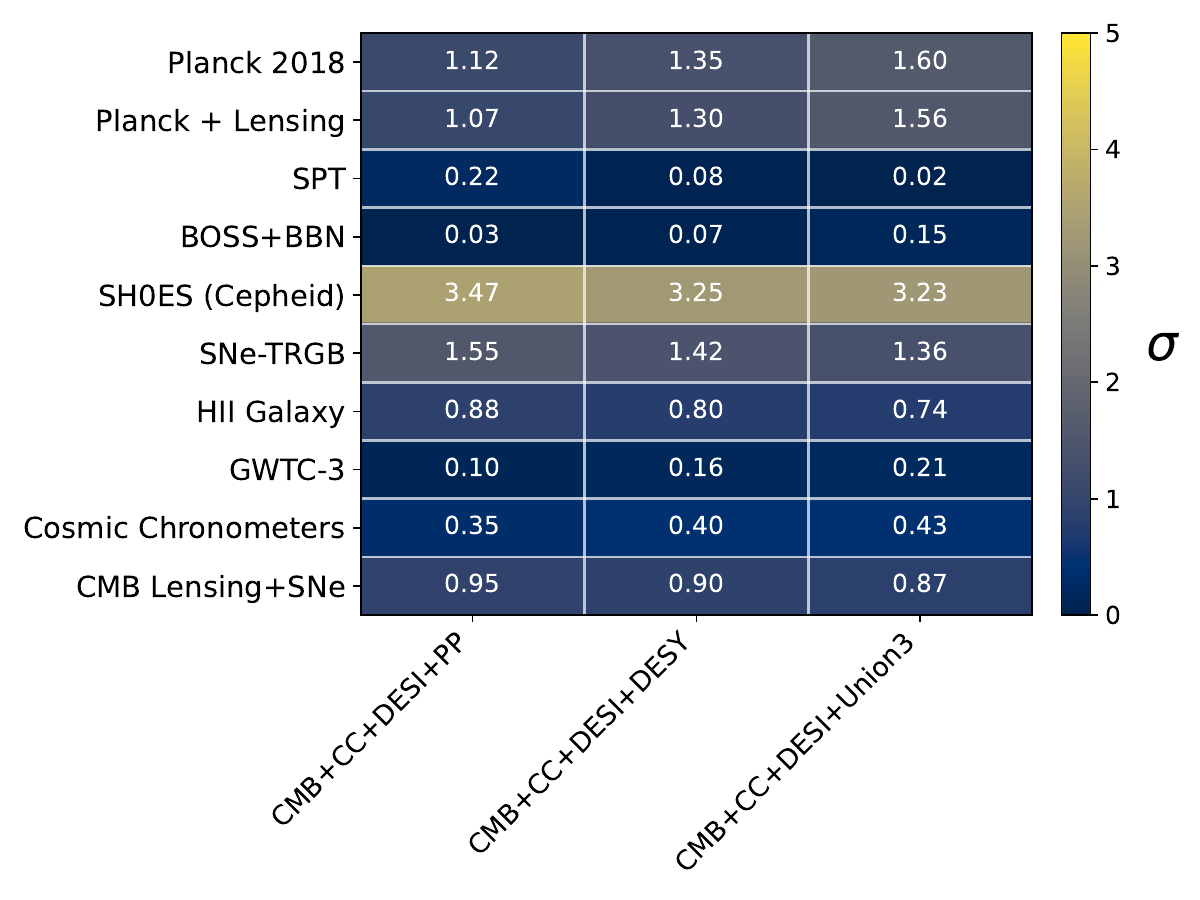}
    \caption{Heat map of the $H_0$ tension between our combinations and various measurements.}
    \label{fig:tension}
\end{figure*}

Fig.~\ref{fig:tension} illustrates the level of tension, expressed in terms of $\sigma$, between the $H_0$ values inferred from the nonminimal $f(Q)$ model and a range of independent measurements reported in the literature. Each cell quantifies the deviation between our model’s best-fit $H_0$ and the corresponding observational value, normalized by the combined uncertainty. 
As expected, the largest discrepancies occur relative to the SH0ES, a tension of approximately $3\sigma$, whereas the model remains broadly consistent with late-time probes such as TRGB, H{\sc II} galaxies, GWTC-3 standard sirens, and cosmic chronometer estimates, all lying within $1$-$2\sigma$. The comparison with ohher probes shows a persistent but reduced tension, reflecting the intermediate position of our model’s $H_0$ between local and CMB-inferred determinations. Overall, the heat map demonstrates that the nonminimal $f(Q)$ model alleviates, though does not entirely resolve, the $H_0$ tension while remaining compatible with the majority of low-redshift measurements.

\section{Machine Leaning Techniques} \label{ML}

This section reviews the primary regression models used for forecasting the Hubble parameter. These models employ distinct underlying assumptions and learning mechanisms, spanning from simple linear relationships to complex nonlinear mappings. Their diversity offers specific advantages for effectively capturing the temporal trends and hidden structures present in the observational data. Machine learning techniques have numerous applications in many fields \cite{patel2026quantifying,libbrecht2015machine,cate2017machine}, underscoring their versatility and potential in cosmological modeling.  In this work, we employ three supervised learning algorithms, Linear Regression, Support Vector Regression (SVR), and Random Forest Regression, to reconstruct the Hubble function directly from observational data. These methods provide complementary perspectives on the underlying expansion history \cite{Elizalde:2021kmo,Arjona:2020kco}. 
\begin{itemize}
\item \textbf{Linear Regression:} Linear Regression (LR) serves as a fundamental and interpretable baseline for reconstructing the Hubble parameter from observational data. By fitting a linear functional relationship between the dependent and independent variables, LR quantifies the overall trend of cosmic expansion in a transparent, analytically tractable manner. Although its performance depends on idealized assumptions, such as the linearity of the underlying relationship and the statistical properties of the errors, it remains a useful first approximation for characterizing large-scale behavior in cosmological datasets. Owing to its low variance, computational efficiency, and straightforward interpretability, LR provides a reliable reference model against which the performance of more flexible nonlinear algorithms can be assessed~\cite{montgomery2021introduction}.

\item \textbf{Support Vector Regression:} Support Vector Regression (SVR) extends the linear approach to more complex regimes by introducing kernel functions that enable nonlinear mappings between input features and the target variable. Depending on the choice of kernel, linear, polynomial, or radial basis function (RBF), SVR can effectively capture both global and local structures in the data. In the cosmological context, this allows the model to identify smooth deviations from linear expansion as well as small-scale variations arising from measurement uncertainties or subtle physical effects. Its strong regularization properties also ensure stability in the presence of noise and limited data, making SVR a powerful method for reconstructing the Hubble function with controlled generalization error~\cite{noble2006support}.

\item \textbf{Random Forest Regression:} Random Forest (RF) Regression is an ensemble-based method that aggregates the predictions of multiple decision trees to reduce variance and enhance predictive robustness. By combining numerous weak learners, the RF algorithm can model highly nonlinear and nonstationary relationships without the need for explicit functional assumptions. This flexibility makes it particularly well-suited for cosmological applications, where the data often exhibit complex dependencies among redshift, distance, and expansion rate. In addition to its resistance to overfitting, RF provides internal measures of feature importance, offering valuable insight into the relative contribution of different cosmological observables~\cite{liaw2002classification}.   
\end{itemize}

The evaluation of predictive performance is carried out using several complementary metrics, each capturing a distinct aspect of model accuracy. The coefficient of determination ($R^2$) quantifies the proportion of variance in the dependent variable explained by the model, serving as a normalized indicator of goodness of fit. Values approaching unity signify that the model accounts for most of the observed variability. It is defined as
\begin{equation}
R^2 = 1 - \frac{\sum_{i=1}^N (z_i - \hat{z}_i)^2}{\sum{i=1}^N (z_i - \bar{z})^2}
\end{equation}
where $z_i$ represents observed values, $\hat{z}_i$ denotes predicted values, and $\bar{z}$ is the mean of observations \citep{draper1998applied}.
To directly quantify prediction errors, we employ the Mean Squared Error (MSE) and the Mean Absolute Error (MAE). The MSE penalizes larger deviations more strongly, making it sensitive to outliers:
\begin{equation}
\text{MSE} = \frac{1}{N} \sum_{i=1}^N (z_i - \hat{z}_i)^2
\end{equation}
while the MAE provides a more intuitive measure of the average magnitude of residuals:
\begin{equation*}
\text{MAE} = \frac{1}{N} \sum_{i=1}^N |z_i - \hat{z}_i|.
\end{equation*}
The lower values of both metrics correspond to a higher predictive accuracy~\cite{hyndman2018forecasting}.

For relative error assessment, we also consider the Mean Absolute Percentage Error (MAPE), which expresses accuracy in percentage terms as
\begin{equation}
\text{MAPE} = \frac{100}{n} \sum_{i=1}^n \left| \frac{z_i - \hat{z}_i}{z_i} \right|.
\end{equation}
This facilitates interpretation and cross-model comparisons \citep{willmott2005advantages}. Together, these metrics provide a comprehensive and balanced framework for evaluating regression models, combining statistical rigor with interpretability in assessing the reconstructed Hubble function.

The observational datasets employed in this analysis correspond to the cosmological models outlined in Section~\ref{data}. For each dataset combination, the inferred values of the Hubble parameter were utilized. These datasets were subsequently analyzed using machine-learning techniques to predict the Hubble parameter from the physical and statistical correlations encoded within the data. The models are trained using 80\% of the dataset to predict observational datasets. After training, the remaining 20\% of the data was used as a testing set to evaluate the predictive performance of each model. This standard split ensures that the evaluation reflects true predictive power rather than overfitting to the training data. The following Table~\ref{ML_Hz_comparison} presents the models' accuracy and error detection rates for the machine learning models used to predict the observational data set.


\begin{figure*}
     \centering
     \includegraphics[scale=0.28]{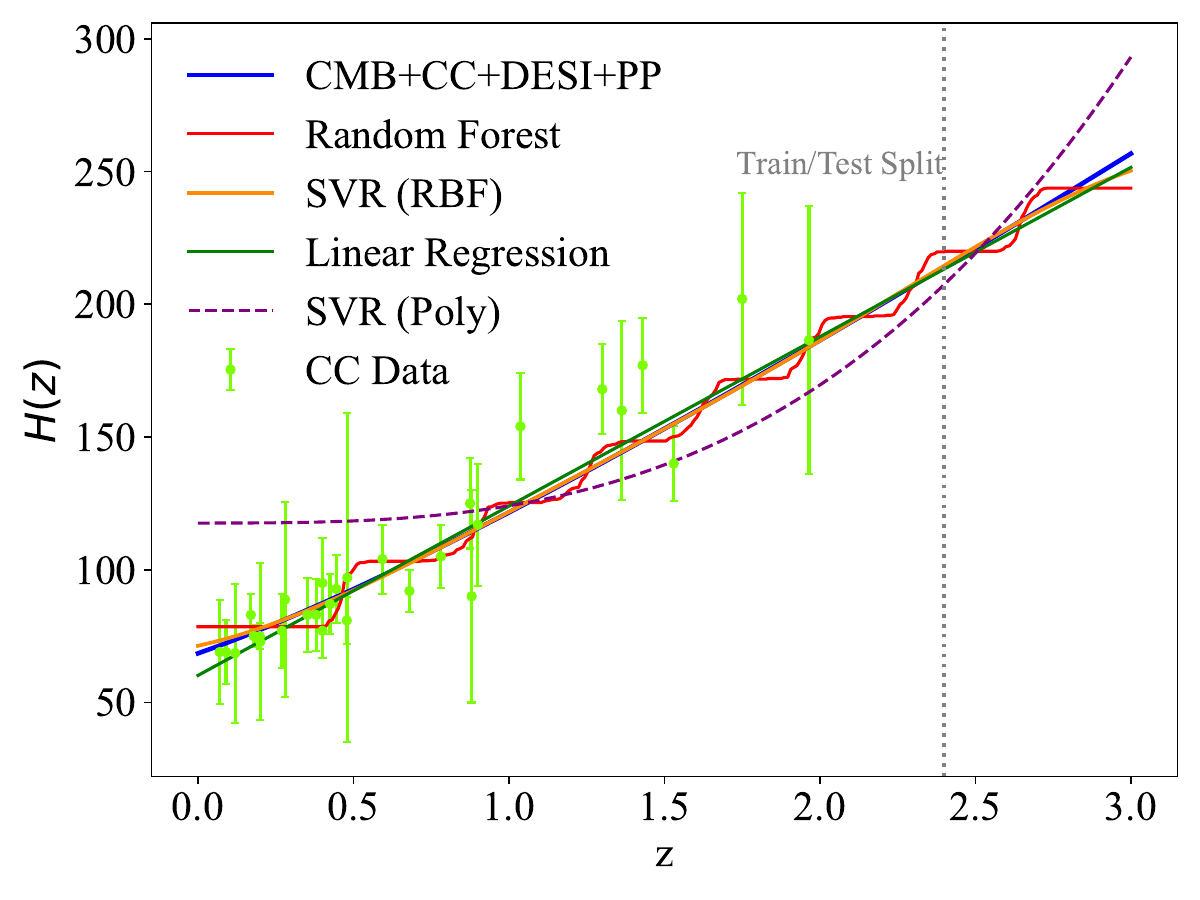}
    \includegraphics[scale =0.28]{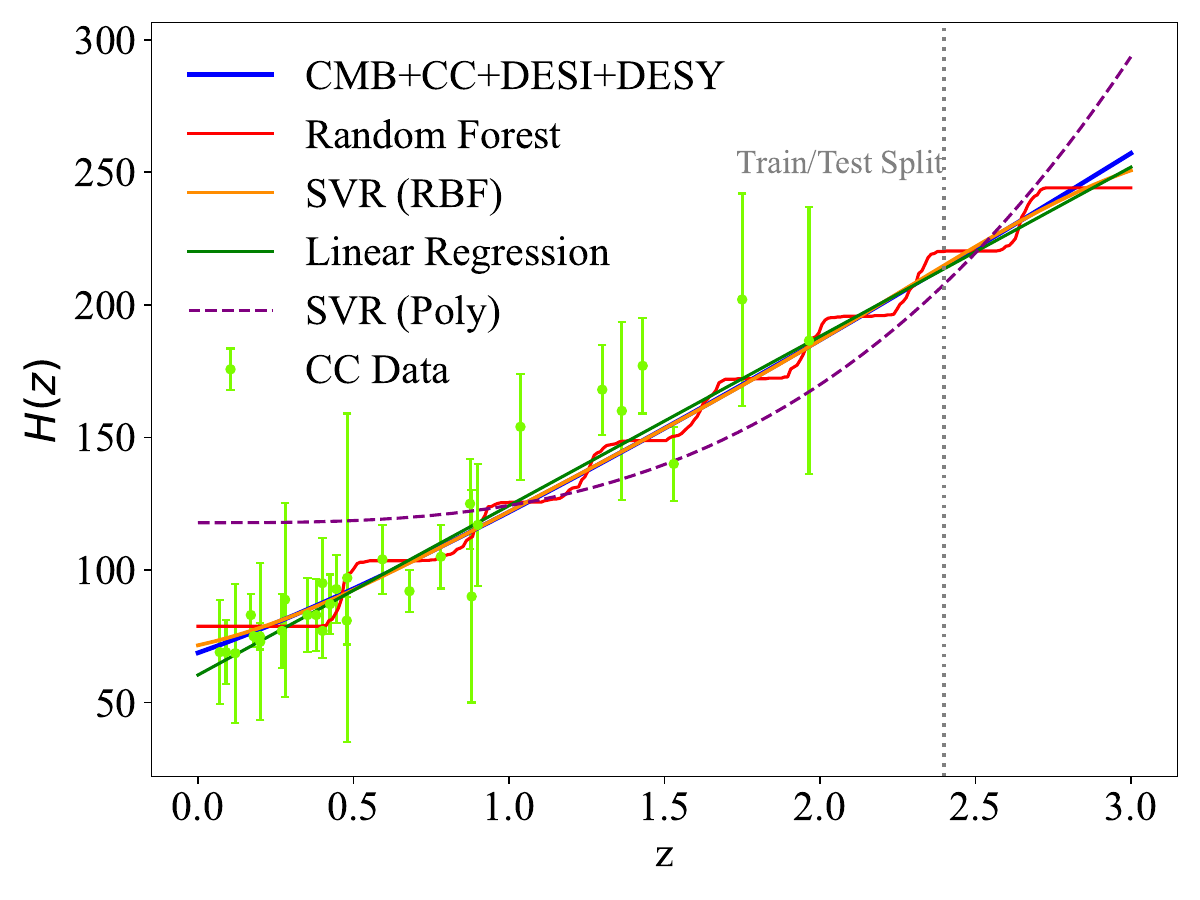}    
    \includegraphics[scale=0.28]{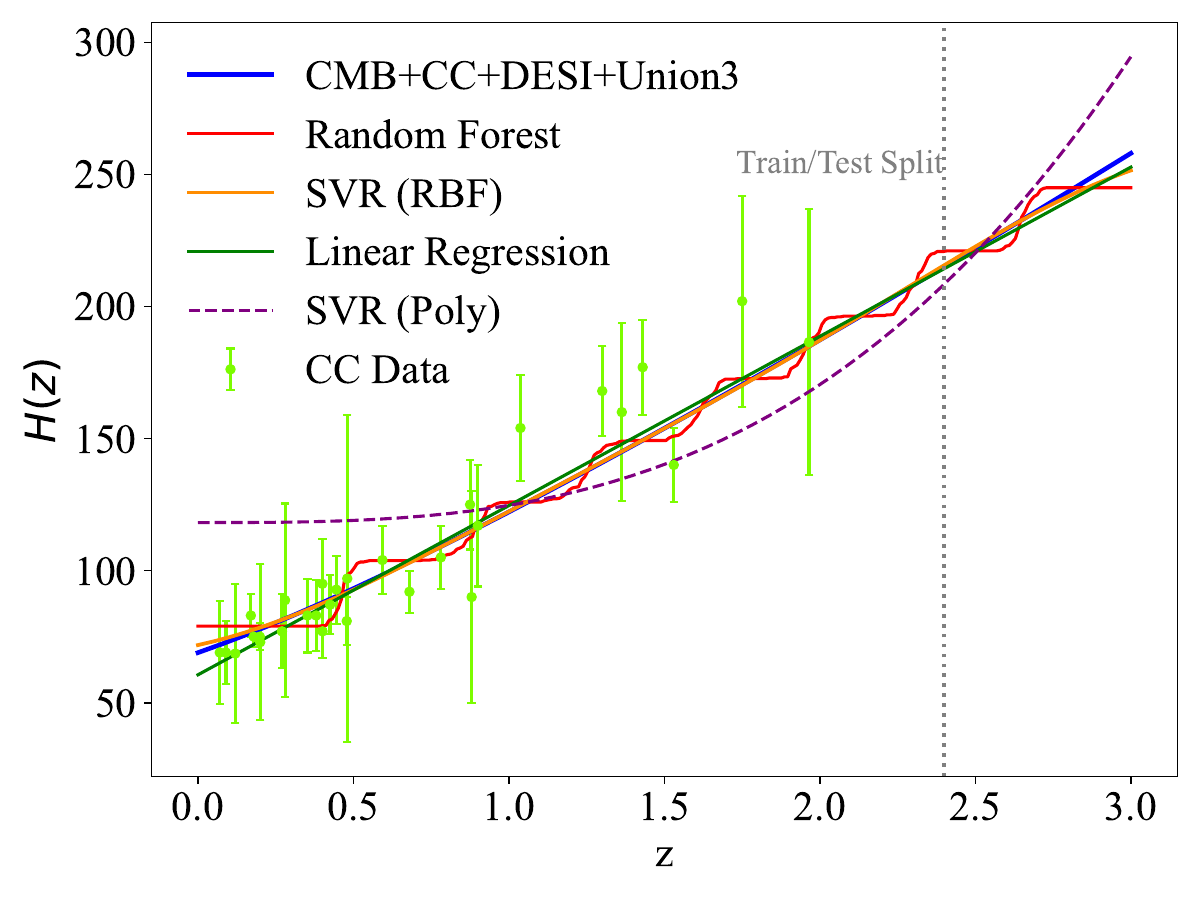}
        \caption{Comparison of theoretical predictions for the Hubble parameter $H(z)$ obtained from different dataset combinations using various machine-learning regression techniques. The figure highlights the ability of these methods to reproduce the observed expansion history across the redshift range.}
    \label{fig:Hmodel_comparison}
\end{figure*}


\begin{table*}[]
\centering
\caption{Comparison of Machine Learning Models on Theoretical $H(z)$}
\label{ML_Hz_comparison}
\renewcommand{\arraystretch}{1.05}
\setlength{\tabcolsep}{3pt}
\begin{tabular}{lccc|ccc|ccc}
\hline
\multirow{2}{*}{Models}
& \multicolumn{3}{c}{CMB+CC+DESI+PP}
& \multicolumn{3}{c}{CMB+CC+DESI+DESY}
& \multicolumn{3}{c}{CMB+CC+DESI+Union3} \\
\cline{2-10}
& $R^2$ & MSE & MAE
& $R^2$ & MSE & MAE
& $R^2$ & MSE & MAE \\
\hline
RF          & 0.9923 & 28.6535 & 4.5726 & 0.9923 & 28.6562 & 4.5721 & 0.9923 & 28.8615 & 4.5884 \\
SVR (RBF)   & 0.9993 &  2.4456 & 0.8453 & 0.9993 &  2.4527 & 0.8460 & 0.9993 &  2.4831 & 0.8518 \\
LR          & 0.9976 &  8.8018 & 2.3506 & 0.9976 &  8.9190 & 2.3660 & 0.9976 &  8.9618 & 2.3717 \\
SVR (Poly)  & 0.8589 & 523.8353 & 19.1693 & 0.8591 & 524.0757 & 19.1766 & 0.8590 & 527.8903 & 19.2457 \\
\hline
\end{tabular}
\end{table*}

\begin{figure*}[]
    \centering
    \begin{subfigure}{0.33\textwidth}
        \includegraphics[width=\textwidth]{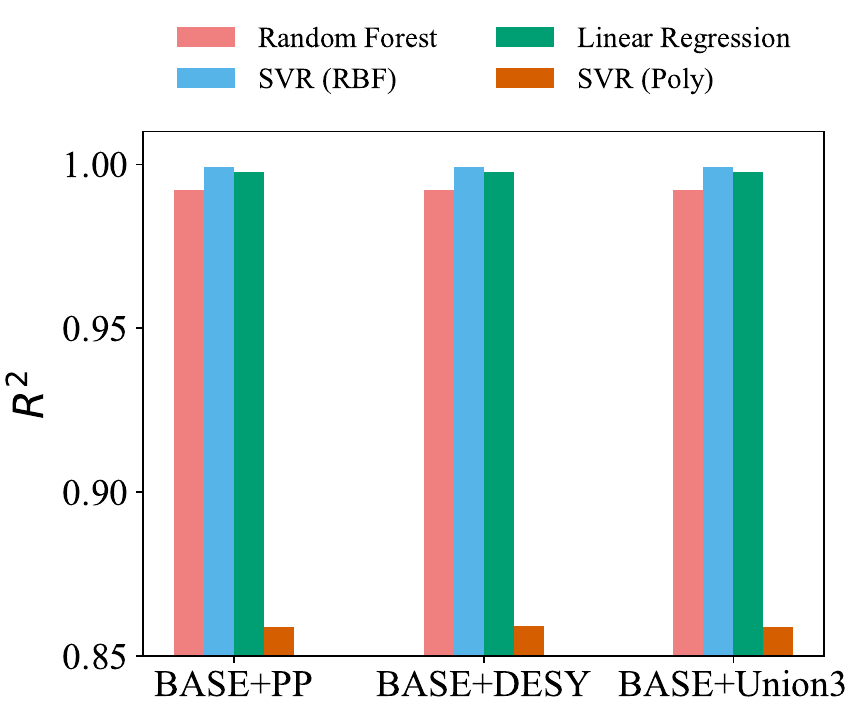}
        \caption{}
        \label{fig:r2}
    \end{subfigure}
\hfill
    \begin{subfigure}{0.33\textwidth}
        \includegraphics[width=\textwidth]{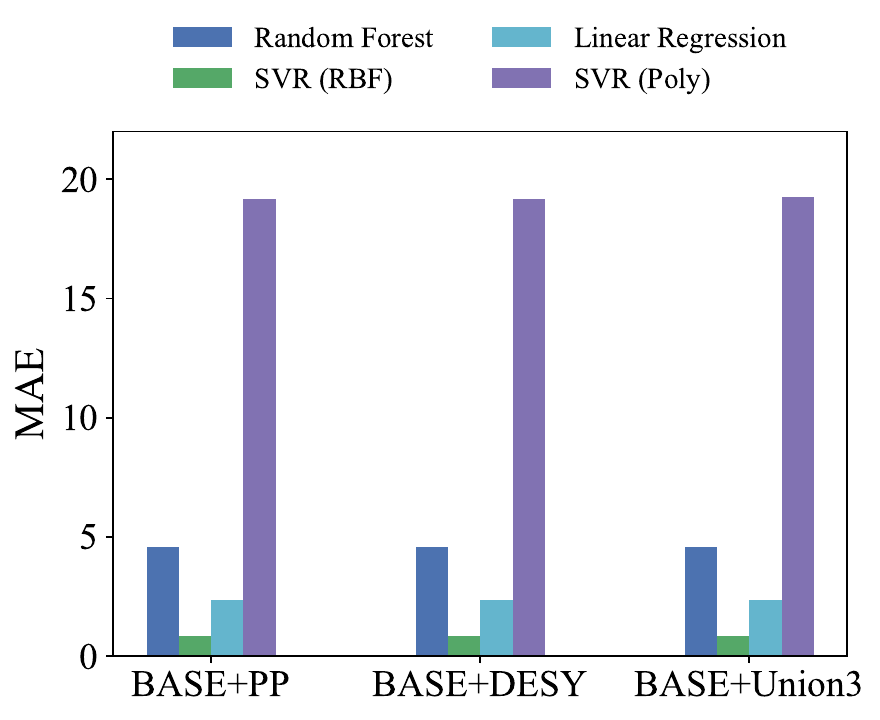}
        \caption{}
        \label{fig:mae}
    \end{subfigure}
\hfill
    \begin{subfigure}{0.33\textwidth}
        \includegraphics[width=\textwidth]{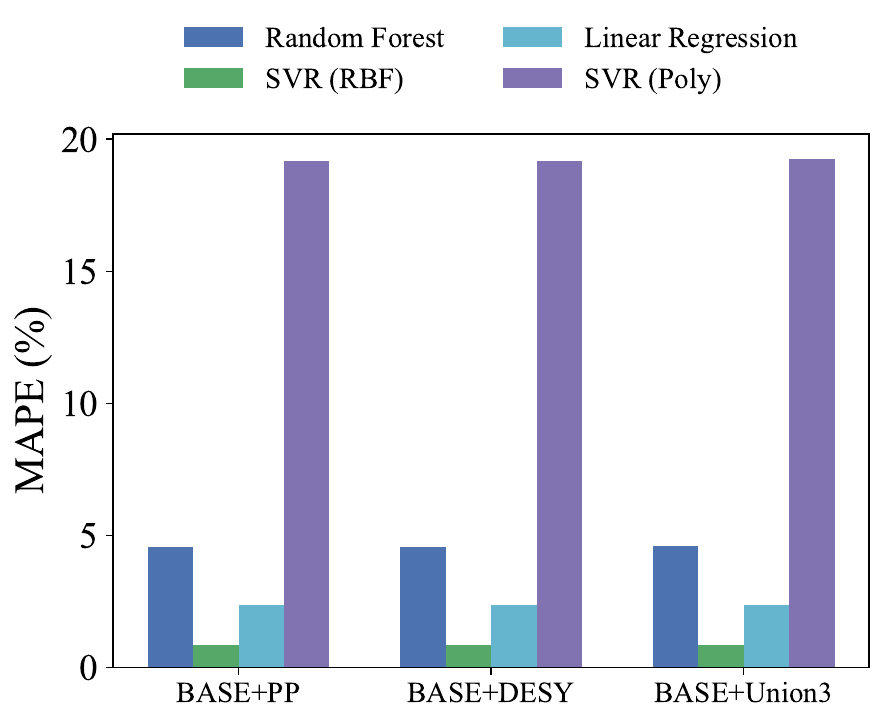}
        \caption{}
        \label{fig:mape}
    \end{subfigure}
    \caption{Performance comparison of different machine-learning models trained on the theoretical $H(z)$ dataset, evaluated using (a) Coefficient of determination ($R^2$), (b) Mean Absolute Error (MAE), and (c) Mean Absolute Percentage Error (MAPE). Here, the BASE data indicates the CMB+CC+DESI dataset.}
    \label{fig:model_comparison}
\end{figure*}

An examination of the $R^2$ reveals a distinct performance hierarchy among each data combination. The SVR employing an RBF kernel demonstrates near-perfect explanatory power, with $R^2$ scores consistently approximating 0.9993 across all data combinations. This indicates an exceptional capacity to capture the variance in the theoretical $H(z)$ data (obtained from each combination of data considered). The linear regression algorithm also exhibits strong performance, maintaining a stable $R^2$ of approximately 0.9976, while random forest yields respectable but comparatively lower values. In contrast, the SVR model with a polynomial kernel lags substantially, suggesting it is ill-suited to the underlying structure of the dataset. A graphical representation of  $R^2$ values from all regression models is shown in Fig. \ref{fig:r2}.

The models' predictive precision is further explained by their error profiles. The SVR (RBF) model distinguishes itself with remarkably low error rates; its MSE and MAE are orders of magnitude smaller than those of other models, underscoring its exceptional accuracy. Linear Regression  produces reliable predictions, though its errors are larger than those of the top performer. Random Forest  follows with a further increase in error magnitude, while the SVR (Poly) model registers the highest errors by a considerable margin, solidifying its position as the least effective approach for the observational data. An analysis of the MAPE, as derived from Table~\ref{ML_Hz_comparison}, reveals that the SVR model employing an RBF kernel exhibits superior performance compared to all other regression models examined. This comparative performance is further illustrated in Figs.~\ref{fig:mae} and \ref{fig:mape}, which present a graphical comparison of both MAE and MAPE values, respectively. 

In conclusion, empirical evidence strongly suggests that the SVR with an RBF kernel is the optimal model for this analysis. Its supremacy is demonstrated through a dual achievement of near-perfect explanatory power and minimal prediction error, a consistency maintained robustly across diverse data combinations. Therefore, the SVR (RBF) model emerges as the most reliable and effective tool for modeling the complex relationships present in the theoretical $H(z)$ data under investigation.

\section{Conclusion} \label{conc}

In this work, we have explored an extension of symmetric teleparallel gravity by introducing a new class of theories in which the nonmetricity scalar $Q$ is coupled nonminimally to the matter Lagrangian within the metric–affine formalism. Similar to standard curvature–matter coupling scenarios, this nonminimal $Q$–matter interaction leads to the nonconservation of the energy–momentum tensor and consequently gives rise to an additional force acting on matter fields. We have further examined an explicit cosmological realization of the theory by adopting specific functional forms for $f_1(Q)$ and $f_2(Q)$, both modeled as power laws. A comprehensive parameter estimation analysis was performed using a broad combination of cosmological datasets, considered in different configurations: (I) CMB + CC + DESI + PP, (II) CMB + CC + DESI + DESY, and (III) CMB + CC + DESI + Union3.

From the contour plots shown in Fig.~\ref{fig:con} and the constraints summarized in Table~\ref{tab:best_fit}, it can be observed that the cosmological parameters exhibit excellent agreement across all dataset combinations. Remarkably, our model reproduces these results while remaining fully consistent with the observational ranges of the corresponding cosmological parameters. The best-fit curves obtained from the MCMC analysis were further compared with the observational datasets, incorporating uncertainties up to the $2\sigma$ confidence level. In all cases, the theoretical predictions align well with the observational data.

The evolution of the Universe begins in an accelerating phase, with the deceleration parameter $q(z)$, taking negative values of the order $q \sim -0.80$. As the cosmic evolution proceeds, the acceleration dynamics exhibit a smooth and nontrivial behavior, with $q$ gradually decreasing over time. In the asymptotic future, the Universe approaches an exponentially accelerating de Sitter phase characterized by $q \rightarrow -1$, a behavior that remains largely independent of the model parameters.

Another key diagnostic, the effective EoS parameter $w(z)$, has also been examined. The present-day values lie within the range $-1 < w_0 < -\tfrac{1}{3}$, confirming a quintessence-like behavior of dark energy in this framework. In addition, we have discussed the existing cosmological tensions and assessed how our model addresses them. A heat map analysis was performed to visualize the statistical significance of these tensions, particularly in relation to $H_0$. It is found that, in most cases, our theoretical predictions lie between the direct (late-time) and indirect (early-time) measurements, effectively acting as a bridge that partially alleviates the $H_0$ discrepancies. Although the present work focuses exclusively on background cosmology, the nonminimal matter--geometry coupling is expected to modify the evolution of matter density perturbations and consequently the growth of cosmic structures. In particular, additional coupling-dependent contributions may arise in the effective growth equation at Newtonian sub-horizon scales. A detailed investigation of these effects and their observational consequences will be presented in future work.

Furthermore, we performed a statistical comparison between our models and the standard $\Lambda$CDM cosmology. The minimum $\chi^2$ values obtained for different datasets provide a quantitative measure of how well each model reproduces the corresponding observations. In order to further investigate the robustness of the cosmological evolution predicted by the nonminimal \(f(Q)\) gravity model, we complement the standard observational analysis with a machine learning based reconstruction study. Using the best-fit parameters obtained from the combined observational datasets, we numerically solve the modified cosmological field equations and reconstruct the corresponding Hubble expansion history \(H(z)\). The generated theoretical \(H(z)\) dataset is then employed as the input for different machine learning algorithms in order to examine their ability to learn, reconstruct, and predict the expansion dynamics associated with the modified gravity framework.

The motivation behind this analysis is to provide an additional data-driven consistency test of the reconstructed cosmological evolution and to explore the effectiveness of machine learning techniques in capturing the nontrivial behavior emerging from modified gravity cosmologies. Therefore, the machine learning analysis should be viewed as a complementary reconstruction and validation framework built directly upon the cosmological solutions of the present nonminimal \(f(Q)\) gravity model.

Nevertheless, our proposed model demonstrates the additional ability to mitigate existing discrepancies, particularly those related to current cosmological tensions, thereby offering a promising direction for further exploration. Motivated by its success in one of the most pressing issues in modern cosmology, future studies may extend this framework to address other open problems in general relativity and late-time cosmic acceleration. Continued research along these lines will help clarify whether nonmetricity-based gravity can provide a compelling and self-consistent alternative to the concordance $\Lambda$CDM paradigm.

\begin{acknowledgments}

\noindent S. A. acknowledges the Japan Society for the Promotion of Science (JSPS) for providing a postdoctoral fellowship during 2024-2026 (JSPS ID No.: P24318). This work of S.A. is also supported by the JSPS KAKENHI grant (Number: 24KF0229). We would like to thank the anonymous reviewer and editor for comments and suggestions that helped us to significantly improve our work. \end{acknowledgments}

\section*{Data Availability}
Data sharing is not applicable to this article as no datasets were generated during the current study.
\bibliography{reference}

\end{document}